\documentclass[twocolumn,showpacs,pre,floatfix,amsmath,amssymb,notitlepage]{revtex4-1}
\usepackage{epsfig}
\usepackage{subfigure}
\usepackage{amsmath}
\usepackage{color}
\usepackage{amssymb}
\usepackage{setspace}
\usepackage{graphicx}
\usepackage{dcolumn}
\usepackage{bm}
\usepackage{times}
\usepackage{enumerate}
\usepackage{footnote}
\input epsf

\graphicspath{{figures/}}

\begin{document}

\author{Digesh Chitrakar}
\affiliation{Department of Mathematics, Trinity College, Hartford, CT 06106, USA}

\author{Per Sebastian Skardal}
\email{persebastian.skardal@trincoll.edu} 
\affiliation{Department of Mathematics, Trinity College, Hartford, CT 06106, USA}

\title{Chaos in Nonlinear Random Walks with Non-Monotonic Transition Probabilities}

\begin{abstract}
Random walks serve as important tools for studying complex network structures, yet their dynamics in cases where transition probabilities are not static remain under explored and poorly understood. Here we study nonlinear random walks that occur when transition probabilities depend on the state of the system. We show that when these transition probabilities are non-monotonic, i.e., are not uniformly biased towards the most densely or sparsely populated nodes, but rather direct random walkers with more nuance, chaotic dynamics emerge. Using multiple transition probability functions and a range of networks with different connectivity properties, we demonstrate that this phenomenon is generic. Thus, when such non-monotonic properties are key ingredients in nonlinear transport applications complicated and unpredictable behaviors may result.
\end{abstract}


\maketitle

\section{Introduction}\label{sec:01}

Random walks have long served as a powerful tool for studying and understanding the structural properties of complex networks due to the wealth of information they provide with remarkably simple dynamics~\cite{Noh2002PRL,Masuda2017PR}. Examples of the widespread utility of using random walks and diffusion dynamics in the context of complex networks include Google's PageRank centrality~\cite{Brin1998,Page1999,Gleich2015SIAM}, network search and exploration~\cite{GomezGardenes2008PRE,Sinatra2011PRE,Battiston2016NJP}, modeling transport, diffusion, and movement processes~\cite{Gorenflo2002,Nicosia2017PRL,Cencetti2018PRE}, detecting communities and other network structures~\cite{Rosvall2008PNAS,Asllani2018PRL}, and finding geometric properties and topological embeddings for dimensionality reduction~\cite{Coifman2005PNAS}. The Markovian, i.e., memory-less, property of a random walk coupled with the static nature of transition probabilities (or transition rates) yields a linear dynamical system where, assuming the relatively mild condition of a network structure being primitive, the dynamics are guaranteed to converge to a unique, globally-attracting fixed point or stationary distribution~\cite{Durrett,MacCluer2000SIAM}. However, in an effort to generalize these conditions to better model more realistic and complex phenomenon, the static nature of transition probabilities may be relaxed, giving rise to nonlinear random walks~\cite{Kolokoltsov2010,Frank2013,Carletti2020PRR}. In particular, by focusing on the case of discrete-time where nonlinearity arises from the the dependence of transition probabilities on the current system state in a manner that biases random walkers towards or away from nodes that are heavily populated, a rich landscape of nonlinear phenomena has been observed, including period-doubling bifurcations, multistability, and quasi-periodic dynamics~\cite{Skardal2019JNS,Skardal2020PRE}.

In this Letter we extend this paradigm to consider nonlinear random walks with non-monotonic transition probabilities. In particular, while the nonlinear random walks with monotonic transition probabilities studied in Refs.~\cite{Skardal2019JNS,Skardal2020PRE} operate on the modeling premise that random walkers are preferentially biased towards nodes with either proportionally more or fewer random walkers present, allowing for non-monotonic transition probabilities allows this bias to be generalized. For instance, depending on the shape of the function that defines the transition probabilities, random walkers might be biased strongly towards {\it both} nodes where very many and very few random walkers are present, while being biased away from moderately populated nodes. Alternatively, random walkers might be biased more strongly towards nodes where a moderate number of random walkers are present, while being biased away from nodes that are both heavily and sparsely populated. In general, such non-monotonic transition probabilities may arise in nonlinear transport applications where the decision-making process that informs movement is nuanced. Examples may include personal travel, where an individual or family may prefer to visit either very remote (e.g., camping) or very densely populated (e.g., large cities) vacation destinations, and finances and investments, where individuals may seek to invest in or choose to purchase commodities from institutions that are neither exceedingly large nor small. Here we explore the dynamics that emerge from nonlinear random walks with non-monotonic transition probabilities and show that they give rise to chaotic dynamics that are not present when transition probabilities are monotonic. We show that this phenomenon is widespread by exploring multiple choices for functions that define the transition probabilities of a system as well as a range of different network topologies. Our results suggest that in nonlinear transport processes where non-monotonicity in transition probabilities or transition rates is a key ingredient, the resulting behavior may be extremely complicated, making prediction and forecasting difficult even with very accurate system information.

The remainder of this paper is organized as follows. In Sec.~\ref{sec:02} we present the governing equations and present our main results, illustrating the emergence of chaotic dynamics in nonlinear random walks with non-monotonic transition probabilities. In Sec.~\ref{sec:03} we explore dynamics on different network topologies, using both a minimal model and a large real-world network. In Sec.~\ref{sec:04} we present an initial exploration of mean return times. In Sec.~\ref{sec:05} we conclude with a discussion of our results.

\section{Governing Equations and Chaotic Dynamics}\label{sec:02}

\begin{figure*}[t]
\centering
\epsfig{file =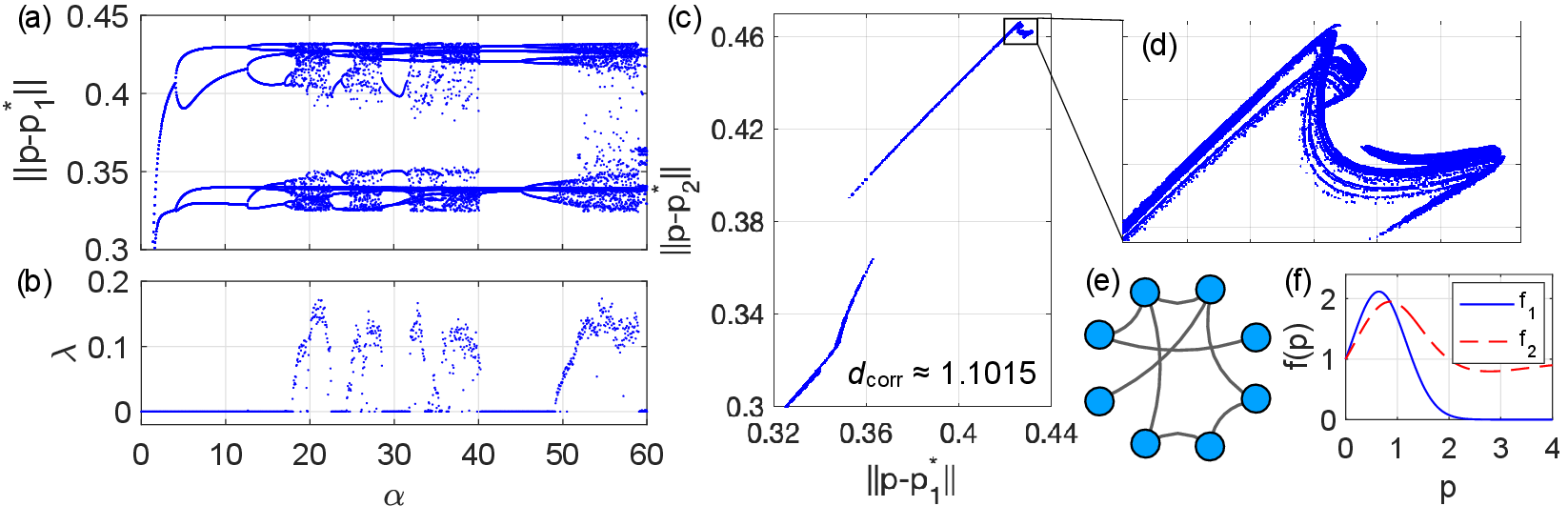, clip =,width=1.00\linewidth }
\caption{{\it Chaos in nonlinear random walks.} For a small network of size $N=8$ and the transition probability function $f_1$ given in Eq.~(\ref{eq:03}) using $z=1.288/N$, (a) the bifurcation diagram using the quantity $\|\bm{p}-\bm{p}_{1}^*\|$ and (b) the largest Lyapunov exponent $\lambda$ as $\alpha$ is increased from $0$ to $60$. (c) For the chaotic parameter value $\alpha = 52.2$ the strange attractor plotted in the $(\|\bm{p}-\bm{p}_1^*\|,\|\bm{p}-\bm{p}_2^*\|)$ plane with (d) a zoomed-in view of the fractal structure. (e) The network structure used for panels (a)--(d) and (f) illustration of the transition functions $f_1$ and $f_2$ using $\alpha =3$ and $z=1.288$ and $2$, respectively.} \label{fig1}
\end{figure*}

We focus our attention on the case of discrete-time nonlinear random walks on complex networks. In the thermodynamic limit of infinitely-many random walkers, the dynamics on a network with $N$ nodes evolves according to
\begin{align}
p_i(t+1) = \sum_{j=1}^N\pi_{ij}(\bm{p}(t))p_{j}(t),\label{eq:01}
\end{align}
where $p_i(t)$ is the fraction of random walkers present at node $i$ at time $t$ for $i=1,\dots,N$ and $\pi_{ij}(\bm{p}(t))$ denotes the transition probability of a random walker moving from node $j$ to node $i$ at time $t$. Importantly, the transition probabilities in the transition matrix $\Pi(\bm{p}(t))$ depend on the current system state, denoted by the vector $\bm{p}(t) = [p_1(t),\dots,p_N(t)]^T$, yielding a nonlinear dynamical system. Since the entries of $\Pi(\bm{p}(t))$ may also be viewed as conditional probabilities, the columns must sum to one, conserving the total probability in the vectors $\bm{p}(t)$ under the dynamics of Eq.~(\ref{eq:01}), which may also be written in vector form $\bm{p}(t+1)=\Pi(\bm{p}(t))\bm{p}(t)$. The most natural formulation for the transition probabilities $\pi_{ij}(\bm{p}(t))$ is that they depend proportionally on the set of states $p_i(t)$ at destination nodes $i$ that stem from node $j$, yielding
\begin{align}
\pi_{ij}(\bm{p}(t))=\frac{a_{ij}f(p_i)}{\sum_{l=1}^Na_{lj}f(p_l)},\label{eq:02}
\end{align}
where $a_{ij}$ is the entry in the adjacency matrix $A$ corresponding to the existence/strength of the link $j\to i$ and $f$ is a function that maps non-negative values to positive values, $f:[0,\infty)\to(0,\infty)$. Unless otherwise noted, here we focus on the case undirected, binary networks such that $a_{ij},a_{ji}=1$ if nodes $i$ and $j$ are connected, and otherwise $a_{ij},a_{ji}=0$, although these conditions may be easily relaxed to explore dynamics on weighted or directed networks.

Previous work investigating the dynamics of Eqs.~(\ref{eq:01}) and (\ref{eq:02}) focused on the monotonic function $f(p)=\text{exp}(\alpha p)$, where $\alpha$ is a biasing parameter that resulted in random walkers being preferentially moved towards heavily or sparsely populated nodes for positive and negative values of $\alpha$, respectively. However, to relax this condition and allow for more nuanced biasing properties we consider the following functions:
\begin{align}
f_1(p) &= \text{exp}[\alpha (p/z)-\alpha(p/z)^2],~~\text{and}\label{eq:03}\\
f_2(p) &= \text{exp}[\alpha \tanh(p/z)-\alpha\tanh((p/z)^2)].\label{eq:04}
\end{align}
For positive $\alpha$ the function $f_1(p)$ increases from $f(0)=1$, reaches a local maximum at $p=z/2$, then decreases with $f_1(p)\to0^+$ as $p\to\infty$. Similarly, $f_2(p)$ initially increases, reaches a local maximum (roughly at $p=0.43z$), and begins to decrease, but then reaches a local minimum before leveling off with $f_2(p)\to1^-$ as $p\to\infty$. [Both $f_1(p)$ and $f_2(p)$ are plotted in Fig.~\ref{fig1}(f) using $\alpha = 3$ and $z=1.288$ and $2$ for $f_1$ and $f_2$, respectively.] Noting that $\alpha$ controls the range of both $f_1$ and $f_2$ while $z$ controls the location of the local maxima and minima, we treat $\alpha$ as the primary bifurcation parameter that will be varied in this work, and use $z$ as a scaling parameter that primarily is chosen to suit the size of the underlying network topology. Note also that setting $\alpha=0$ yields $f(p)=0$, which in turn yields $\pi_{ij}=a_{ij}/k_j$ in Eq.~(\ref{eq:02}), i.e., the classical static, unbiased random walk. In this case, provided that the network structure is primitive (i.e., the adjacency matrix is irreducible and aperiodic) the dynamics converge to a unique stationary state given by the normalized degree vector, $\bm{p}^*=\bm{k}/\sum_{i}k_i$ where $\bm{k}=[k_1,\dots,k_N]^T$ and $k_i=\sum_{j=1}a_{ij}$. In contrast to these simple linear dynamics, transition probabilities defined by the function $f$ in Eq.~(\ref{eq:03}) yields richer dynamics.

We begin by illustrating our main result using a small network of size $N=8$ with the function $f_1$ given in Eq.~(\ref{eq:03}). Setting $z=1.288/N$, we plot in Fig.~\ref{fig1}(a) the bifurcation diagram using the quantity $\|\bm{p}(t)-\bm{p}_1^*\|$ as $\alpha$ is increased from $0$ to $60$. For sufficiently small $\alpha$ the dynamics converge to a fixed point near $\bm{p}\approx\bm{p}_1^*$, but soon undergo a period-doubling bifurcation at $\alpha\approx 0.6$. Beyond this point we observe a cascade of further period-doubling bifurcations, eventually giving way to chaotic dynamics with subsequent bands of higher-order periodic and chaotic behavior, as is typical in other generic maps displaying chaotic dynamics~\cite{Ott2002}. To confirm that the dynamics are in fact chaotic, we plot in Fig.~\ref{fig1}(b) the largest Lyapunov exponent $\lambda$, which agrees nicely with the bifurcation diagram. We note here that the conservation of probability (or more generally, the conservation of the sum of $\bm{p}(t)$) in Eqs.~(\ref{eq:01}) and (\ref{eq:02}) implies that the dynamics lie on the simplex characterized by $\sum_{i=1}^Np_i(t)=1$, $p_i(t)\ge0$ for $i=1,\dots,N$. Moreover, the Jacobian matrix for Eq.~(\ref{eq:01}) always has at least one eigenvalue given precisely by $\lambda = 1$, making this simplex a center manifold that is marginally stable in the transverse direction away from the simplex. Since perturbations in this direction neither grow or decay, this guarantees that at least one Lyapunov exponent will always be zero, so the largest Lyapunov exponent will be $\lambda=0$ for non-chaotic (e.g., periodic and quasi-periodic) dynamics and $\lambda>0$ for chaos. Next we probe the chaotic dynamics further, setting $\alpha = 52.2$ and plotting the chaotic attractor in the $(\|\bm{p}(t)-\bm{p}_1^*\|,\|\bm{p}(t)-\bm{p}_2^*\|)$ plane, where $\bm{p}_2^*=\bm{1}/N$ is the normalized constant vector, in Fig.~\ref{fig1}(c). The stretching and folding that is typically present in strange attractors can be more easily identified by zooming in on, e.g., the upper-right hand side of the attractor, as shown in Fig.~\ref{fig1}(d). To confirm the fractal nature of the structure we perform a fractal analysis that reveals a correlation dimension~\cite{Ott2002} given by the non-integer value of $d_{\text{corr}}\approx1.1015$. In Fig.~\ref{fig1}(e) we illustrate the network structure used in Figs.~\ref{fig1}(a)--(d).

\begin{figure}[t]
\centering
\epsfig{file =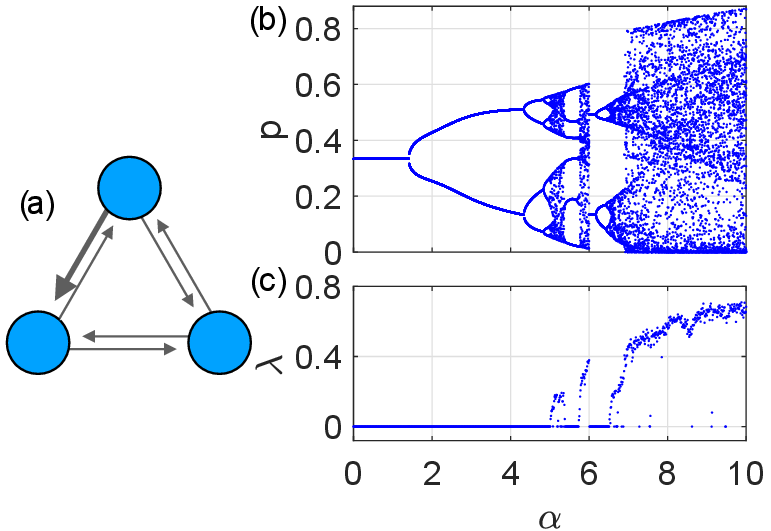, clip =,width=0.93\linewidth }
\caption{{\it A minimal system for chaos.} (a) A three-node network with a single directed link strengthened to twice that of the other directed links. For this minimal network (a) the bifurcation diagram using the quantity $\|\bm{p}-\bm{p}_{1}^*\|$ and (b) the largest Lyapunov exponent $\lambda$ as $\alpha$ is increased from $0$ to $20$ using $f_1$ with $z=1.288$.} \label{fig2}
\end{figure}

Before continuing on, we pause to discuss the implications of these new findings. While the applicability of classical random walks (i.e., those with static transition probabilities, either biased or not) cannot be overstated, the relaxation of static transition probabilities is a key ingredient in modeling transport dynamics when decision-making is more nuanced. The results presented above show that when non-monotonicity is a feature in such a process, then not only may the the system not converge to a simple fixed point or periodic orbit of low order, but possibly a higher-order periodic orbit or even a chaotic state. Moreover, in the case of a chaotic state we lose long-term predictability of the system behavior, posing a hurdle for forecasting beyond the very short-term, even with very accurate system information.

\section{Minimal Model and Varying Network Topology}\label{sec:03}

We now demonstrate that the chaotic dynamics illustrated in the example above are generic in the sense that they occur for multiple transition probability functions as well as over a wide range of network topologies. We begin with the fundamental task of finding a minimal system that exhibits chaos. Note that, since we do not allow for self-links, i.e., random walkers may not remain at the same node from one time step to the next, $N=3$ is the smallest network with non-trivial dynamics. In fact, chaos can be observed by breaking the symmetry of a complete 3-node network by increasing the strength of one directed link in the network, as illustrated in Fig.~\ref{fig2}(a), where the thick arrow indicated a directed link with strength two while the rest of the directed links have strength one. Using $f_1$ and $z=1.288/N$, we plot in Fig.~\ref{fig2}(b) the bifurcation diagram for the probability $p_i(t)$ describing the upper-most node in Fig.~\ref{fig2}(c) and in Fig.~\ref{fig2}(c) we plot the corresponding largest Lyapunov exponent. Thus, with only three nodes we observe hallmarks of chaos such as period-doubling bifurcations, bands of periodicity and chaos, and positive Lyapunov exponents. We note here that we found the symmetry breaking described above to be necessary to destabilize the stationary state fixed point, although this does not rule out the possibility for multistability with a chaotic orbit, although we found no such structures in our exploration.

\begin{figure}[t]
\centering
\epsfig{file =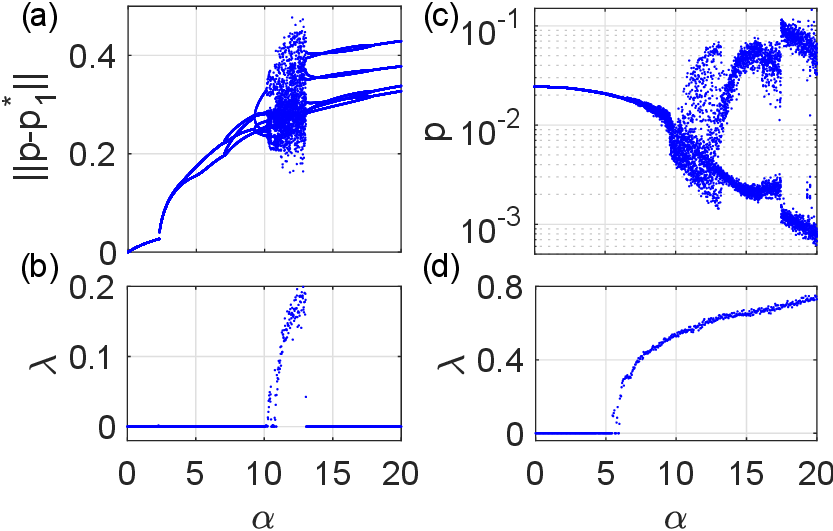, clip =,width=1.00\linewidth }
\caption{{\it Varying transition probability functions and network topologies.} For the probability transition function $f_2$ given in Eq.~(\ref{eq:04}) with $z=2/N$, the bifurcation diagram using the quantity $\|\bm{p}-\bm{p}_{1}^*\|$ with network topologies given by (a) the network depicted in Fig.~\ref{fig1}(e) and (c) the 2002 US airport network. (b),(d) The largest Lyapunov exponent $\lambda$ as $\alpha$ is increased from $0$ to $20$ for the same networks.} \label{fig3}
\end{figure}

Next we demonstrate that the chaotic dynamics exhibited in the previous examples persists for different probability transition functions and for different network structures. First, using $f_2(p)$ given in Eq.~(\ref{eq:04}) with $z=2/N$ on the same network structure depicted in Fig.~\ref{fig1}(e), we plot the bifurcation diagram and largest Lyapunov exponent as $\alpha$ is increased from $0$ to $20$ in Figs.~\ref{fig3}(a) and (b), respectively. Again, we observe a cascade of period-doubling bifurcations that eventually give way to chaotic dynamics just above $\alpha = 10$ that agree with a positive Lyapunov exponent. We also consider the same dynamics on a much larger network structure generated from real data describing the interactions between the 500 busiest airports in the United States in 2002~\cite{Colizza2007NatPhys}, with nodes representing airports and links existing between airports if they have a direct flight going from one to the other. (Here we consider the undirected, unweighted version of this network.) In addition to being a larger, real-world network, this structure is much more heterogeneous, with several of the $N=500$ nodes having a nodal degree larger than $100$ while many others have only one or a handful of links. Again using $f_2$ with $z=2/N$ we plot the bifurcation diagram and largest Lyapunov exponent as $\alpha$ is increased from $0$ to $20$ in Figs.~\ref{fig3}(c) and (d), respectively. (Here we have plotted the bifurcation diagram using the probability $p_i(t)$ corresponding to the node of largest degree, given by $k_i=145$.) With the increased size, heterogeneity, and overall complexity of the network topology, some signature effects such as period-doubling bifurcations and intermittent bands of periodic and chaotic dynamics are lost, however the largest Lyapunov exponent indicates chaos shortly after $\alpha = 5$.

In addition to the network structures used so far in this Letter, namely the first toy network ($N=8$), the minimal network ($N=3$), and the US airport network ($N=500$), we have simulated the random walk dynamics of Eqs.~(\ref{eq:01}) and (\ref{eq:02}) for a number of number of random networks with different parameters. In the Appendix we present these results, showing several examples for each combination of network structures (i) of size $N=8$ with mean degree $\langle k\rangle=2$, (ii) $N=24$ with $\langle k\rangle=4$, and (iii) $N=100$ with $\langle k\rangle=10$ with both $f_1$ and $f_2$, given by Eqs.~(\ref{eq:03}) and (\ref{eq:04}). We also explore the effect of different choices of non-monotonic functions, specifically presenting results using two additional functions that differ from $f_1$ and $f_2$ in that they begin by decreasing before hitting a pair of local extrema. This larger collection of simulations show qualitatively similar results as those presented above and demonstrate that the chaos we observe is generic in the sense that it occurs over a wide range of network structures and for multiple transition functions.

\section{Mean Return Times}\label{sec:04}

\begin{figure}[t]
\centering
\epsfig{file =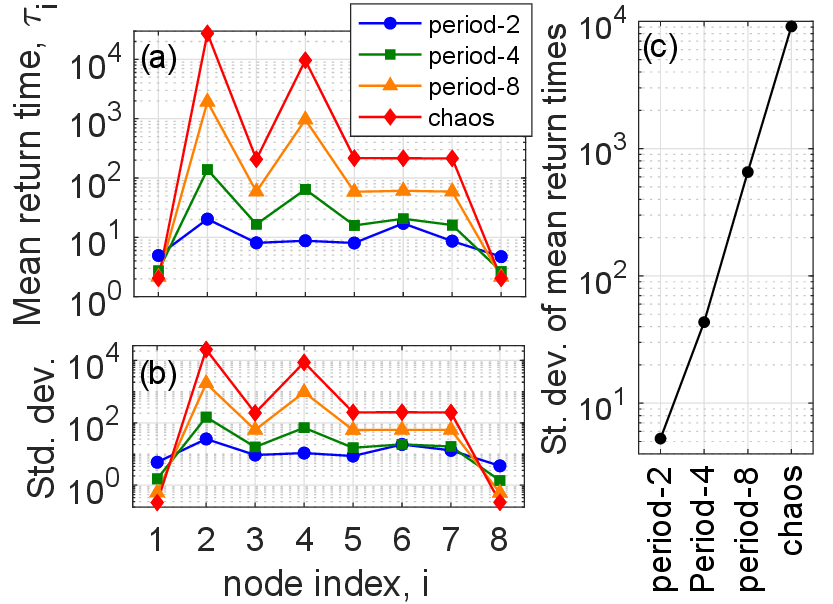, clip =,width=1.00\linewidth }
\caption{{\it Effect of dynamics on mean return times.} (a),(b) Mean and standard deviation of return times observed in simulations with period-2, period-4, period-8, and chaotic dynamics obtained from choosing $\alpha=3$, $10$, $15$, and $20$, respectively, illustrated by blue circles, green squares, orange triangles, and red diamonds. The function $f_1$ is used with $z=1.288$ on the network illustrated in Fig.~\ref{fig1}(a), with nodes 1--8 in the clockwise direction starting from the top. (c) For each of the four cases, the standard deviation of the mean return times across the network.}\label{fig4}
\end{figure}

Before we conclude, we explore the effect that the complicated dynamics that emerge in nonlinear random walks with non-monotonic transition functions have on characteristic return times, i.e., the typical time it takes for a random walker to return to a particular node. In particular, returning to the toy network illustrated in Fig.~\ref{fig1}(e) $f_1$ with $z=1.288$, we consider the four choices $\alpha=3$, $10$, $15$, and $20$, yielding period-2, period-four, period-8, and chaotic dynamics. In each scenario, we simulate the trajectory throughout the network of a particular random walk (using $10^6$ timesteps) to calculate the mean and standard deviation of return times for each node, which we plot in Figs.~\ref{fig4}(a) and (b), using blue circles, green squares, orange triangles, and red diamonds for the four different kinds of dynamics. In particular, as the underlying dynamics become more complicated, the return times for different nodes become more heterogeneous, with mean return times ranging between $4.73$ and $2.03\times10^1$ when dynamics are period-2, but $2.04$ and $9.65\times10^3$ when dynamics are chaotic. The overall spread of mean return times across the network is described in Fig.~\ref{fig4}(c), where we plot the standard deviation of mean return times across the network for each of the four cases. While this results constitute only four specific parameter choices, they tend to agree with other exploratory simulations that suggest that more complicated dynamics result in larger heterogeneity in mean return times throughout a network. The topic of return times (as well as hitting times etc.) in nonlinear random walks on networks deserves further attention in future research.

\section{Discussion}\label{sec:05}

In this work we have studied the dynamics that take place in a class of discrete-time nonlinear random walks where transition probabilities depend on the current state of the system. Specifically, we showed that when transition probabilities are non-monotonic, that is, they are defined by a non-monotonic function of the state of the random walk dynamics, complicated behaviors that include chaos emerge. In the case of small networks, the classical signatures of low dimensional chaotic systems such as cascades of period-doubling bifurcations and intermittent bands of chaos and periodicity can be observed. For larger systems, however, these fingerprints become difficult to identify with added complexity, but chaos remains a prominent feature. We also illustrated that these dynamics are generic in the sense that they occur over a wide range of network topologies and with several transition functions with different properties. (In addition to the examples given here, see the Appendix). These results suggest that when the rules governing nonlinear transport throughout a system are made non-static and non-monotonic, for instance, by nuanced decision-making that directs individuals' trajectories, the resulting dynamics may become very rich and complicated, making long-term prediction and forecasting difficult even with very accurate system information.

\bibliographystyle{plain}

\end{document}